# Devising Interactive Spaces: A Rehearsal-Oriented Tool for Creating Responsive Environments for Immersive Theatre

Pavlos Panagiotidis

Mixed Reality Lab, School of Computer Science, University of Nottingham, Nottingham, United Kingdom, pavlos.panagiotidis@nottingham.ac.uk

Paul Tennent

Mixed Reality Lab, School of Computer Science, University of Nottingham, Nottingham, United Kingdom, paul.tennent@nottingham.ac.uk

Jocelyn Spence

Mixed Reality Lab, School of Computer Science, University of Nottingham, Nottingham, United Kingdom, jocelyn.spence4@nottingham.ac.uk

Nils Jaeger

Department of Architecture and Built Environment, Faculty of Engineering, University of Nottingham, Nottingham, United Kingdom, nils.jaeger@nottingham.ac.uk

We present a rehearsal-oriented system for creating responsive built environments during theatre devising workshops. The system connects bespoke sensing modules for gesture, position, and speech recognition to light and sound outputs through a visual no-code programming layer. It was developed, used, and refined across six workshops with eight professional performance-makers, where participants created light-and-sound scores, gesture- and position-triggered scenes, responsive architectures, participatory prototypes, and a multi-room scratch performance. Rather than presenting a production-ready show-control platform, this demo focuses on how sensing and actuation can be made available as compositional materials during early-stage creative experimentation for immersive theatrical compositions. The system is designed to support quick configuration, visible mappings, and in-room testing, allowing performers to experiment with responsive spaces with minimal technical support. We describe the system architecture, its workshop use, and the practical conditions that helped integrate interactive sensing into embodied performance-making.

CCS CONCEPTS • Human-centered computing → Human computer interaction (HCI) → Interaction paradigms → Collaborative interaction • Applied computing → Arts and humanities → Performing arts

**Additional Keywords and Phrases:** Theatre, Live Performance, Devising, Interactive Built Environments, Responsive Environments, Performance-led HCI

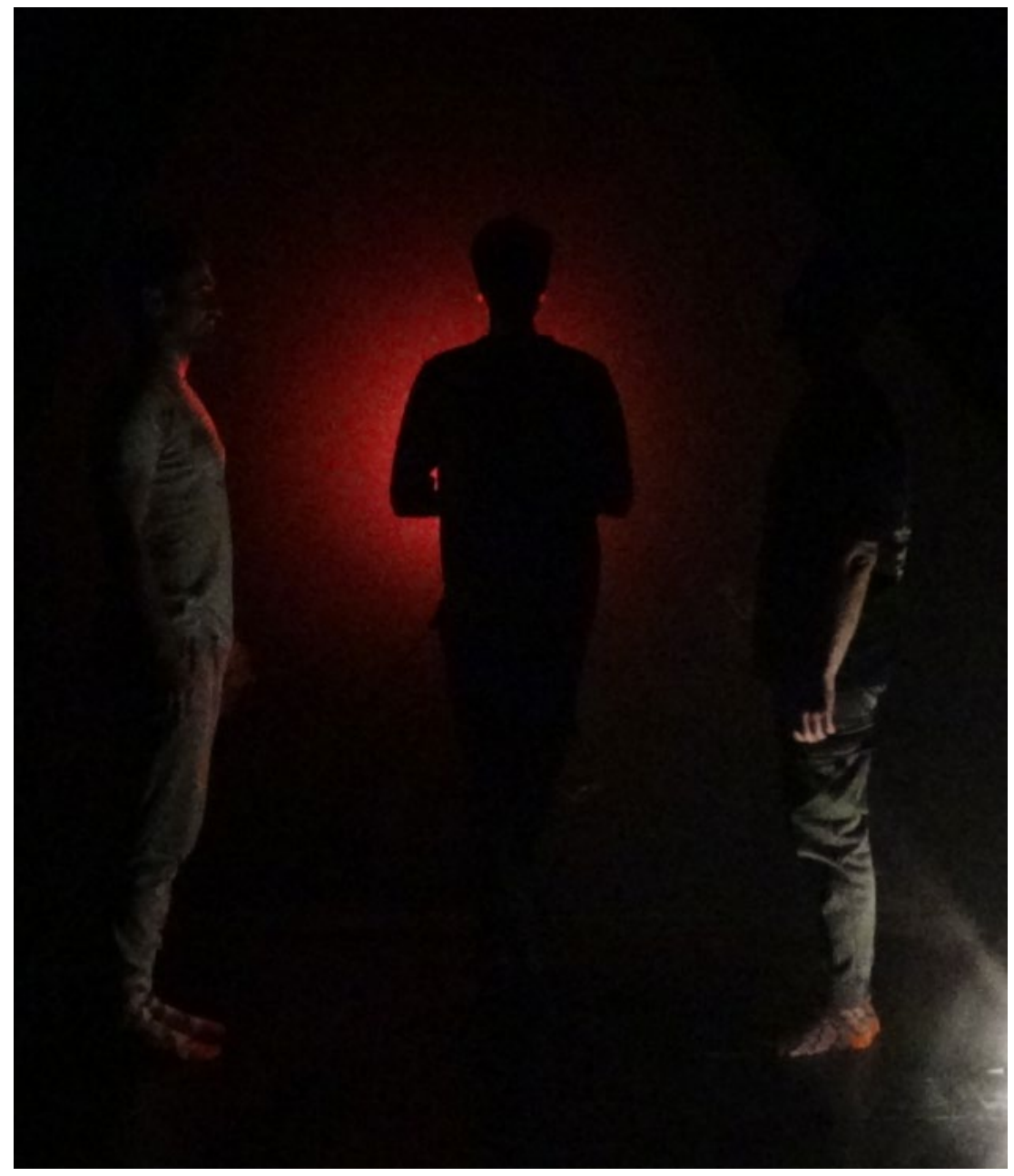

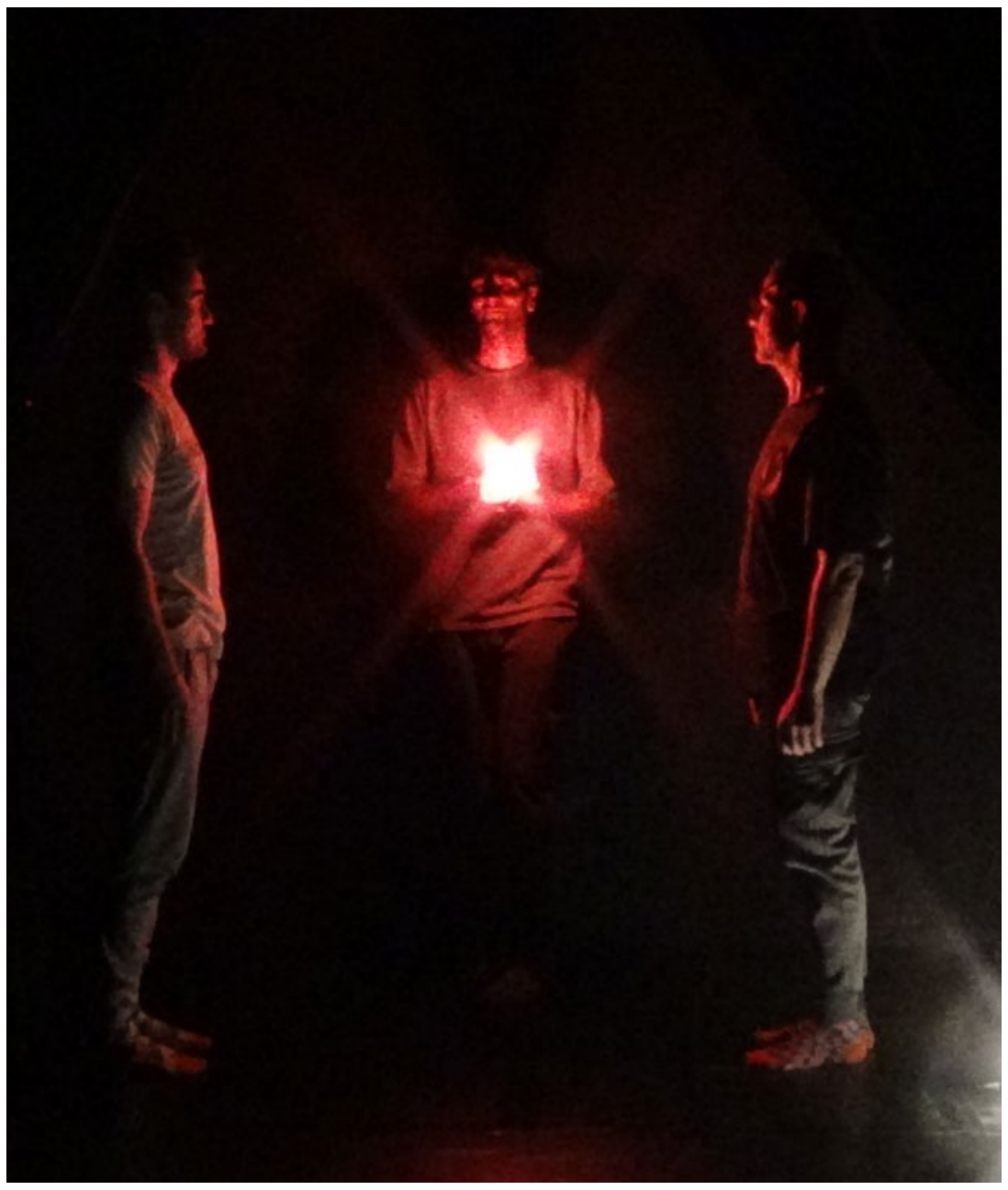

Figure 1: Performers explore movement and spatial relations while configuring and testing a responsive built environment during rehearsal.

## 1. INTRODUCTION

Fiction has long imagined spaces that behave: buildings that sense, anticipate, and sometimes retaliate [2], casting architecture as a responsive presence rather than a static container. This lens is newly useful for contemporary immersive performance, where experience is organised through a world that surrounds the spectator [10], across forms ranging from free exploration to guided trajectories [18]. In such experiences, an interactive built environment can become a dynamic compositional medium, understood less by what it is than by what it does [8]. Although sensing and actuation technologies are becoming increasingly accessible, less is known about how theatre-makers can work with them at rehearsal tempo: quickly configuring, testing, and revising responsive spaces through embodied exploration.

This creates a practical tension. Interactive systems often combine multiple subsystems, require configuration and calibration, and depend on technical oversight to remain stable. Rehearsal, by contrast, depends on rapid embodied trial-and-error exploration, iterative adjustment, and collective judgement in the moment. If responsive architectures are to become theatrical compositional materials, they need to be authorable inside rehearsal through quick experiments, local revisions, and shared control across the creative ensemble, even when technical expertise remains necessary for maintaining and extending the system.

This work builds on performance-led HCI, where rehearsal is treated as a site for generating and testing interaction concepts through relations between bodies, technologies, spaces, and audiences [3, 4]. It also connects to devising and embodied design approaches, where interaction possibilities emerge through situated enactment rather than specification-first workflows [9, 16]. Prior work on interactive performance also recognises breakdown and repair as routine conditions [7, 14]. However, less attention has been given to how theatre-making ensembles can configure and revise responsive environments themselves during early-stage rehearsal, especially when bespoke sensing remains fragile.

We present a rehearsal-oriented system for integrating bespoke sensing into theatre devising workshops. The system connects gesture, position, and speech inputs to light and sound outputs through a visual no-code programming layer. It was developed, used, and refined across six workshops with eight professional performance-makers, who used it to create short interactive scenes, responsive architectures, participatory prototypes, and a multi-room scratch performance. The system was designed to support devising: a collaborative theatre-making process in which dramaturgical meaning is discovered through practical exploration rather than predetermined by a fixed script [11].

Rather than presenting a production-ready show-control platform or the full empirical analysis of the workshops, this paper describes the system and the authoring conditions it was designed to support: quick configuration, visible mappings, and in-room testing. The live demonstration presents a compact version of the rehearsal setup. It shows how sensing inputs, such as body position or a recorded gesture, can be linked through Node-RED to light memories or sound cues and tested immediately through movement. The focus is on the authoring loop used in the workshops: define a trigger, connect it to an environmental response, test, adjust, and repeat.

For theatre-makers and HCI researchers working with immersive performance, the system offers a way to prototype responsive environments before they become fixed technical designs. More specifically, it supports rehearsing interactive conditions: spaces where performers can configure and test mappings, discover behaviours through movement, and decide what role responsive technology should have within a broader composition. Its purpose is therefore not to replace production-grade show-control systems, but to make sensing and actuation available as compositional materials within embodied devising.

## 2. SYSTEM DESCRIPTION

The system is organised as a set of interchangeable "ingredients" (sensing and actuation modules) connected through a visual logic layer, enabling incremental scaling from single-trigger tests to multi-cue compositions.

**Inputs (sensing):** A camera-based pipeline provided two embodied input channels. First, gesture/body tracking was implemented using MediaPipe [6]; participants recorded gestures and defined recognition features based on selected body landmarks and joint-angle relations, producing a stream of recognised gesture events. Second, spatial position tracking used YOLOv8 [17] for person detection from an overhead live video feed. Position data could be logged and visualised during setup to support calibration. A third input channel supported vocal triggering via offline speech-to-text using Vosk [1]. Spoken phrases were transcribed live and matched to keywords, producing discrete trigger events for use in interaction rules. This module was used less due to time constraints and limited transcription fidelity for our participants' language use.

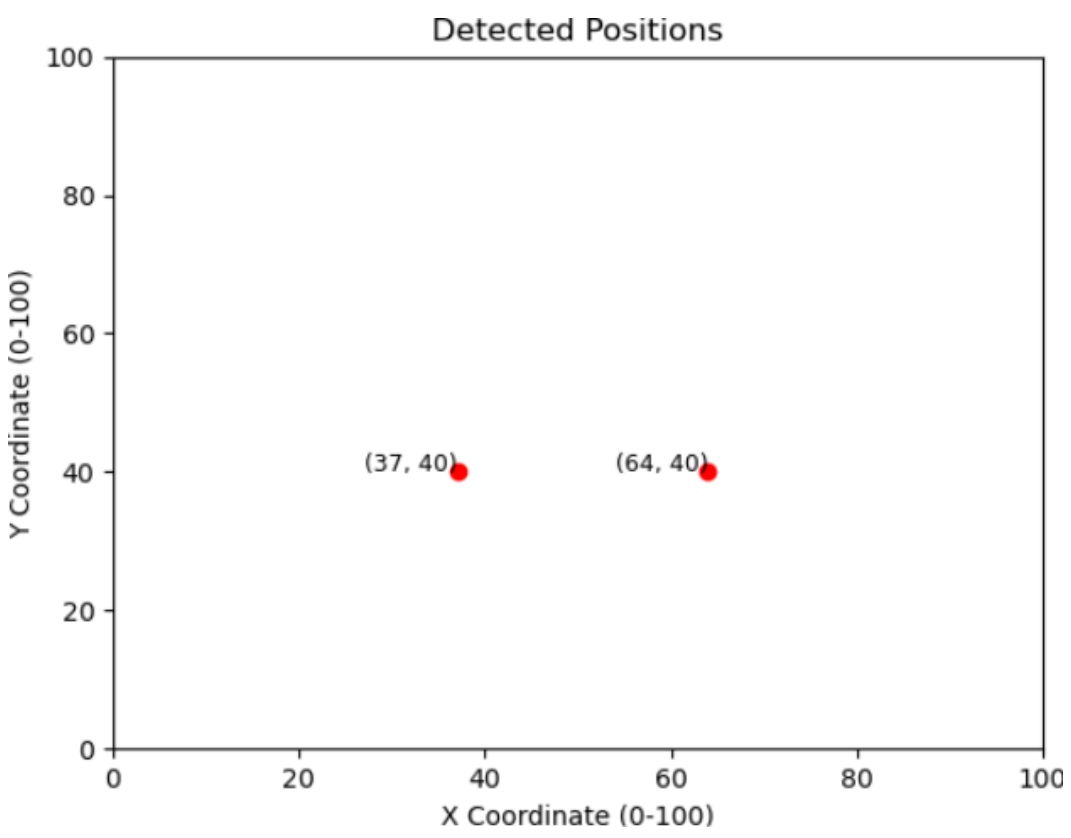

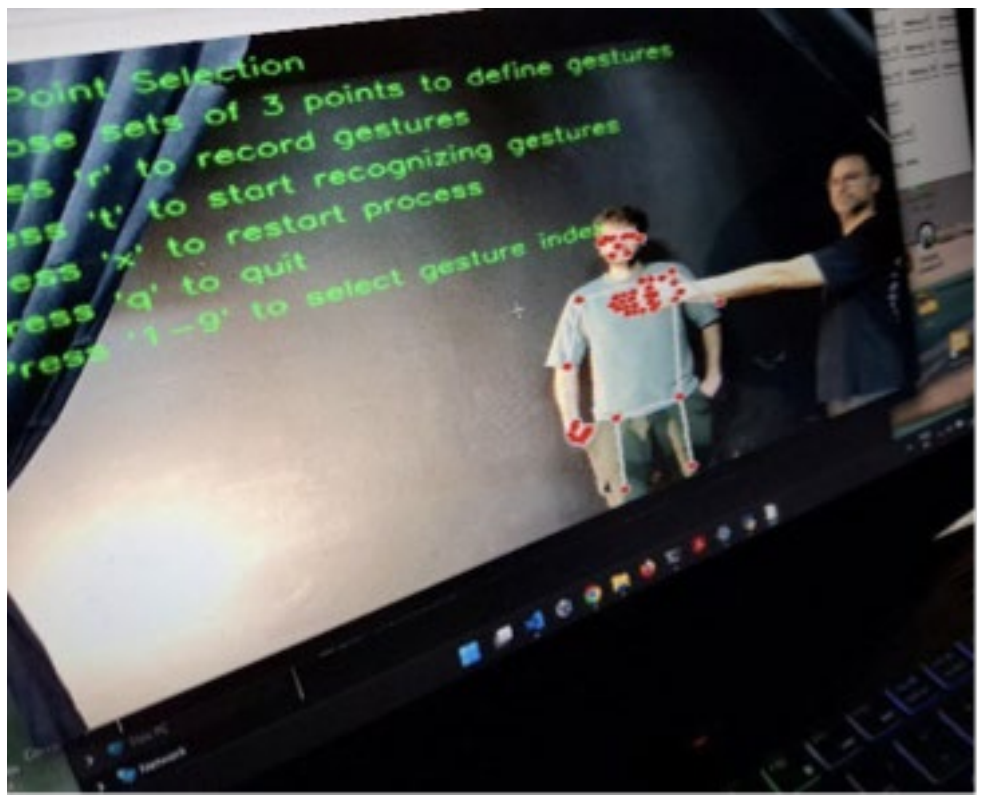

Figure 2: Sensing modules used as input to interaction authoring: (left) camera-based position tracking streams normalised coordinates; (right) gesture/body tracking recognises recorded gestures.

**Logic and integration.** Interaction rules were set in Node-RED [12] as a no-code visual programming environment (Figure 3). In Node-RED, participants created simple conditional flows that read from left-to-right: an input node (gesture, position, or phrase) defined a trigger condition, which then activated an output action node.

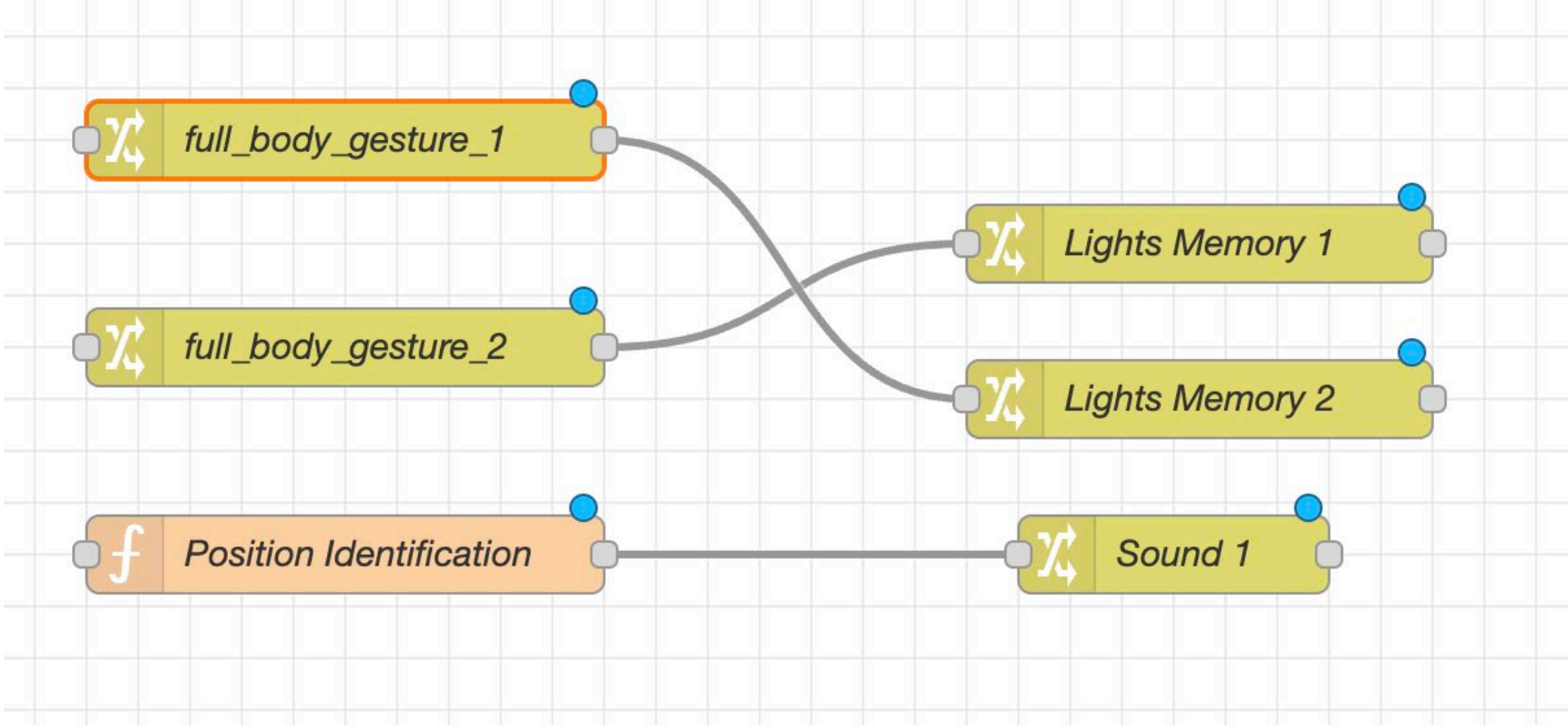

Figure 3: Visual authoring in Node-RED: a left-to-right flow linking an embodied trigger (gesture, position) to a scenographic action (recalling a light “memory” or triggering audio).

**Outputs (actuation):** Lighting was controlled through a custom GUI for Philips Hue [15] smart bulbs, providing direct manipulation of brightness and colour and supporting storage/recall of up to 20 pre-set “light memories” used as scenographic states. Sound output was managed through a custom interface that allowed recording/loading of up to eight audio clips. These output modules could be combined to support mappings such as gesture→light state, position→light or sound cue, and phrase→sound playback. For these workshops, we used bespoke output interfaces rather than a fully featured show-control platform, prioritising simplicity, learnability, and rapid use in rehearsal over advanced cueing functionality.

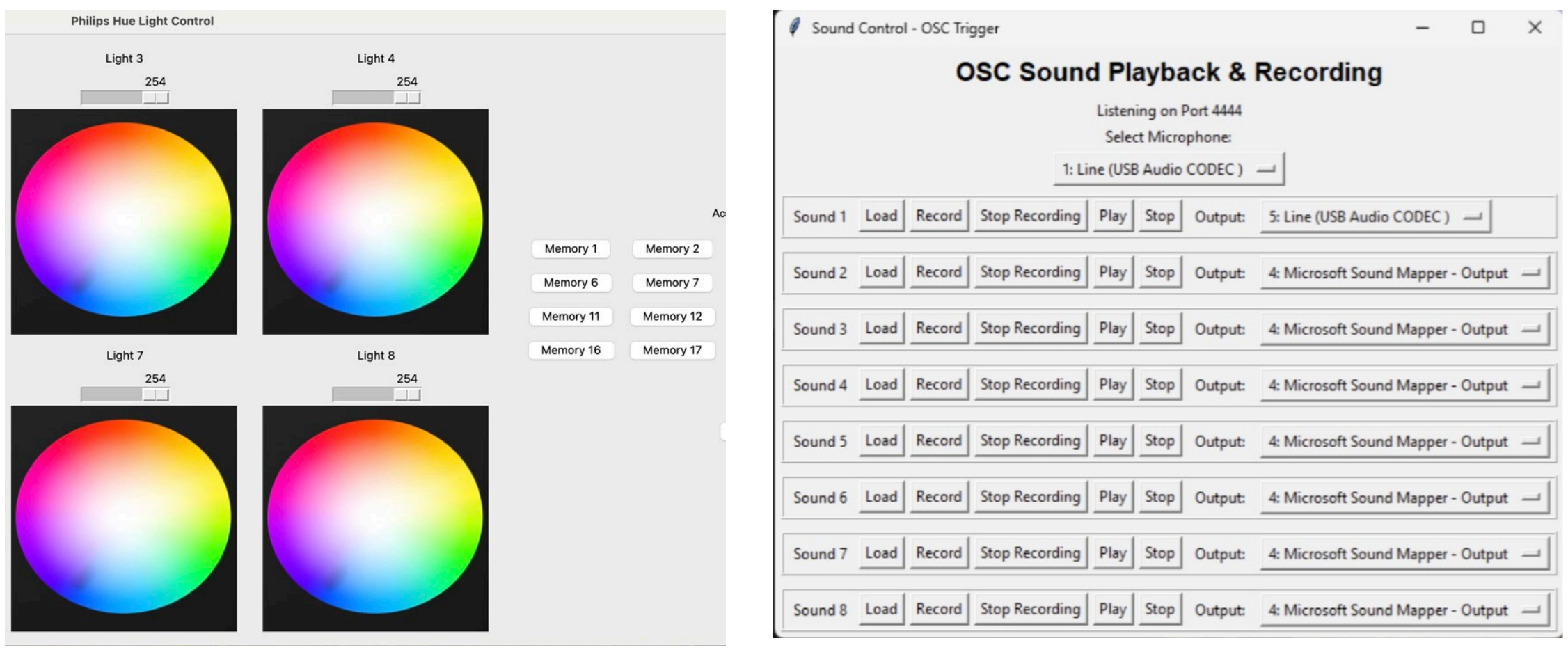

Figure 4: Output control interfaces used during rehearsal: (a) Philips Hue lighting interface with recallable "memories" (preset lighting states); (b) sound interface for recording/loading short audio clips.

## 3. WORKSHOP FORMAT

The study received ethics approval from the School of Computer Science Research Ethics Committee, University of Nottingham (application ID: CS-2023-R72). Participants consented to participation and to the use of audiovisual recordings for research dissemination; where identifiable images are shown, participants provided explicit consent.

Across six devising-oriented workshops, the system was introduced progressively to the eight professional performance-makers so that technical work remained embedded in rehearsal rather than separated into a programming task. We used Viewpoints-informed devising as a shared compositional frame, giving performers a practical vocabulary for working with spatial relations, tempo, movement, and architecture [5]. Each session followed a similar approach: somatic warm-up, small-group structured creative exploration, and internal sharing or testing. This allowed the ensemble to move from simple light and sound control toward more complex responsive environments while keeping the focus on embodied composition.

The workshop sequence began with manual control of light and sound states, establishing output fluency before sensor-based triggering was introduced. Gesture, speech, and position triggers were then added through staged tasks, allowing participants to test simple mappings such as gesture→light, position→sound, or phrase→audio. After the initial familiarisation in the first two workshops, later sessions shifted toward participants' creative autonomy: small groups configured responsive environments using the system, and tested how interactivity became part of performance composition. The system was then used to design responsive architectures for staged scenes or for other performers to inhabit improvisationally. In these cases, the performers entering these spaces did not always know the mappings in advance, but discovered the environment's behaviours through action. The final two workshops extended this toward audience-oriented prototypes and a multi-room scratch performance, where the responsive system was used selectively alongside physical architecture and performer action.

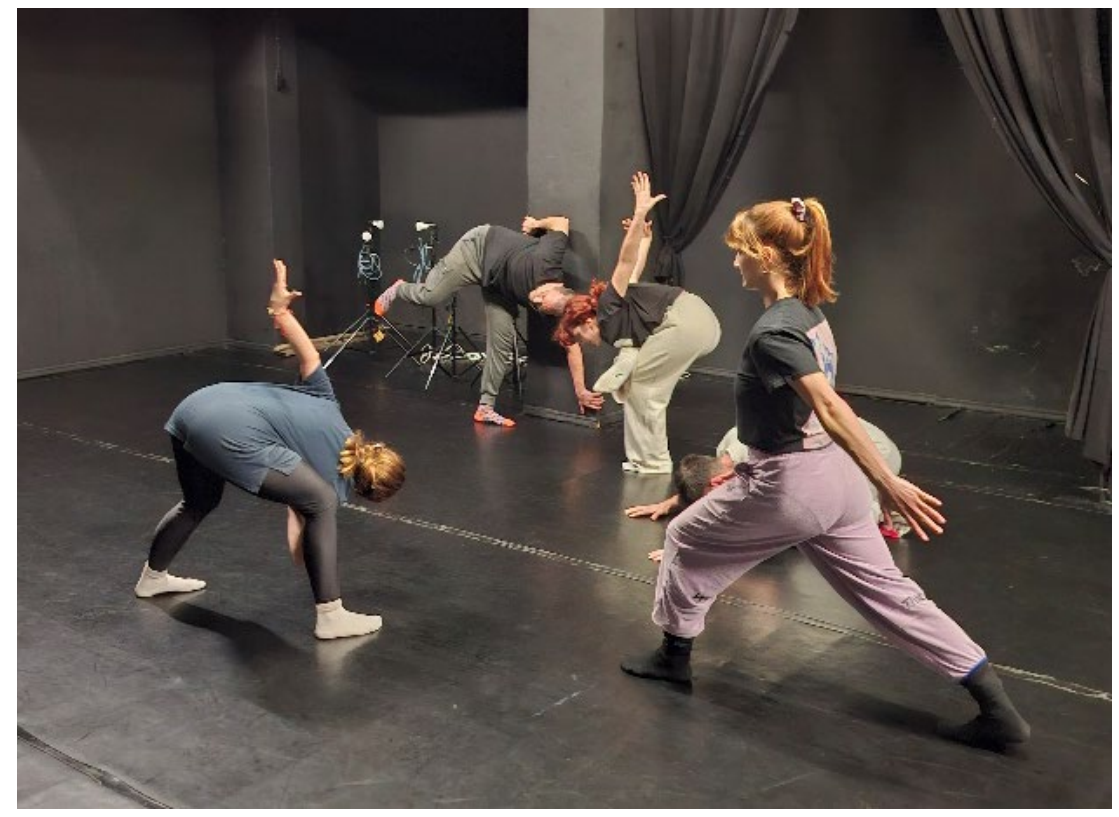
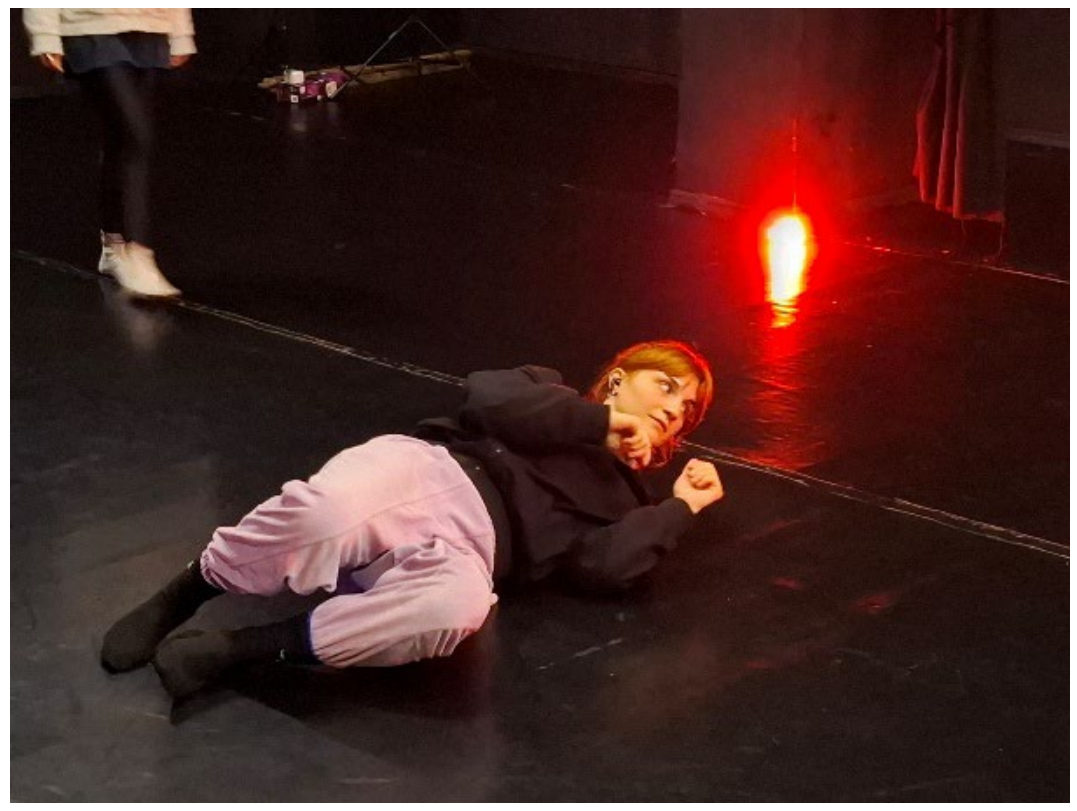

Figure 5: Viewpoints improvisation during rehearsal: (left) open Viewpoints; (right) improvising inside an adaptive responsive space.

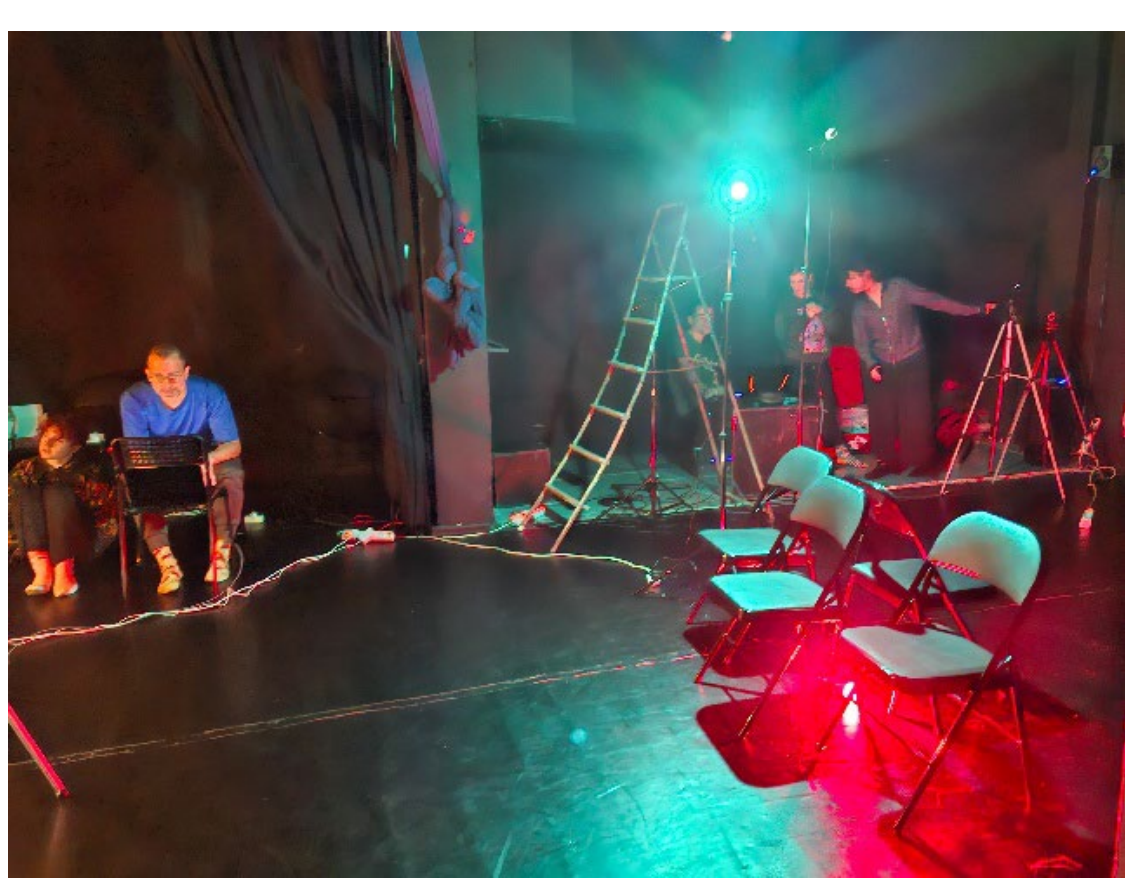
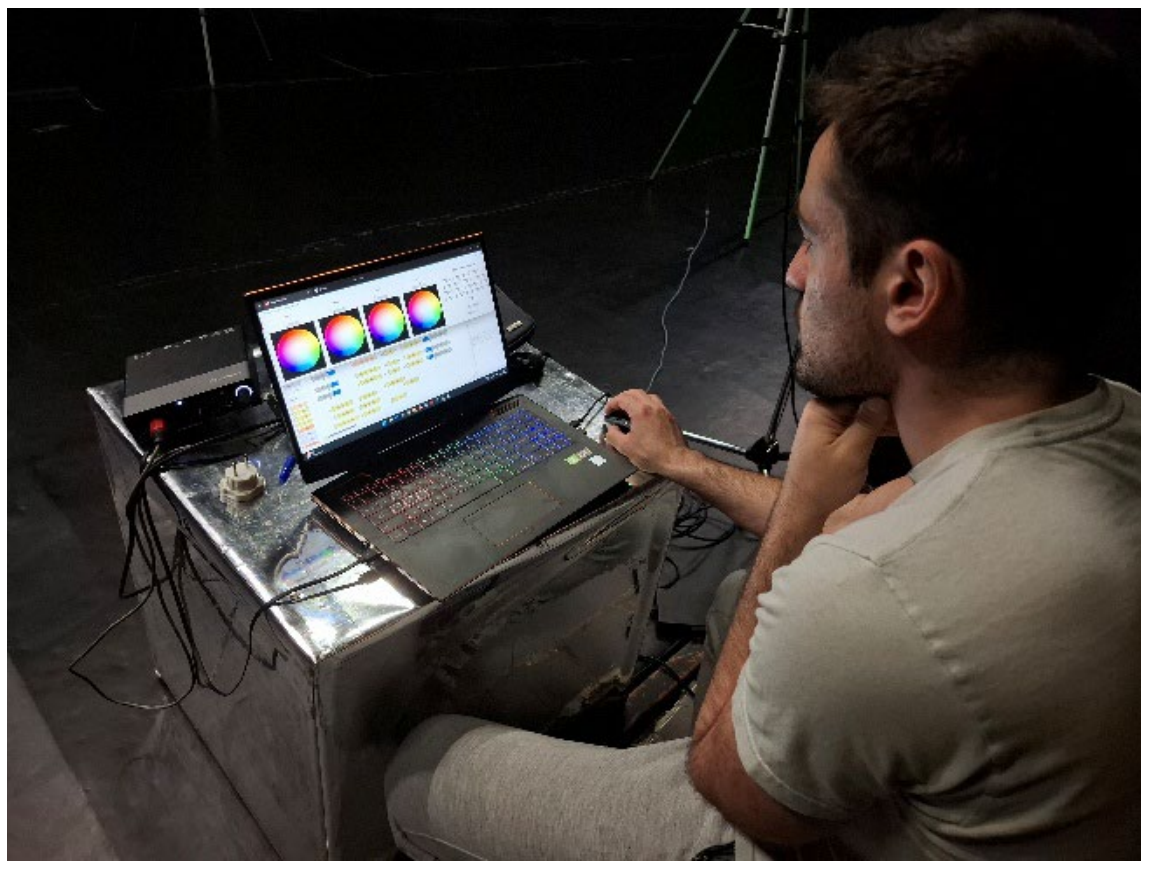

Figure 6: Interaction authoring setup in rehearsal.

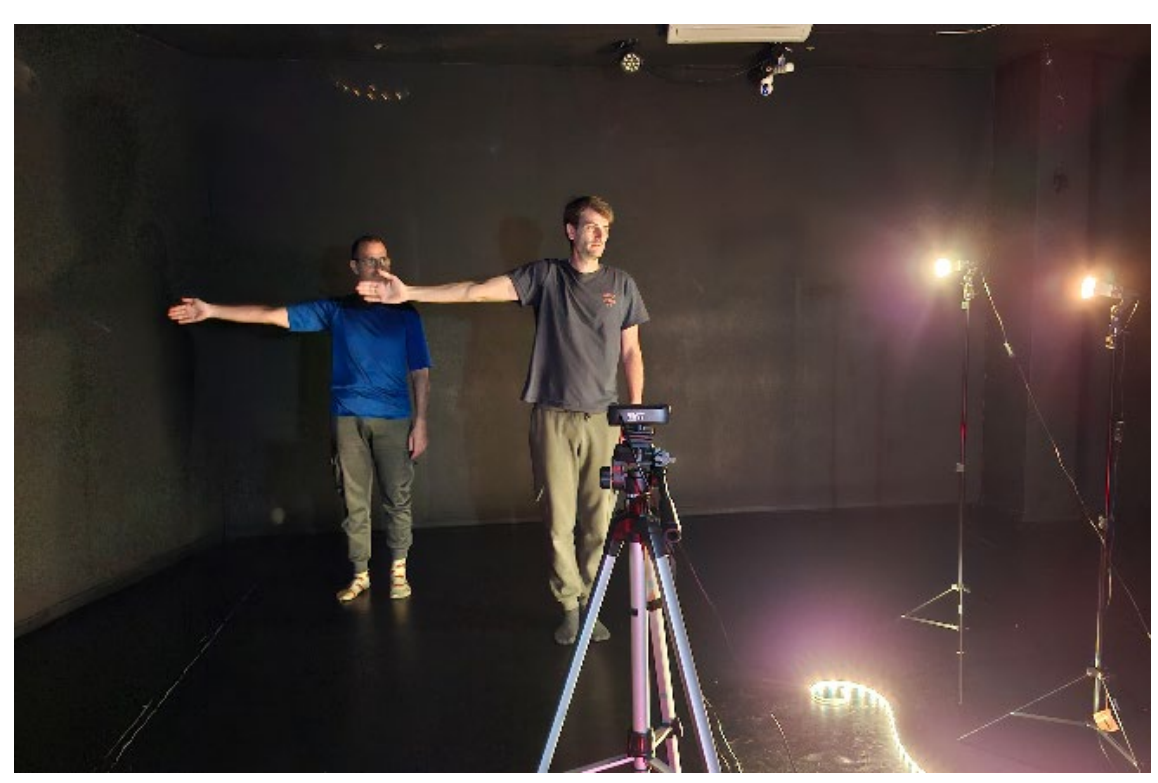
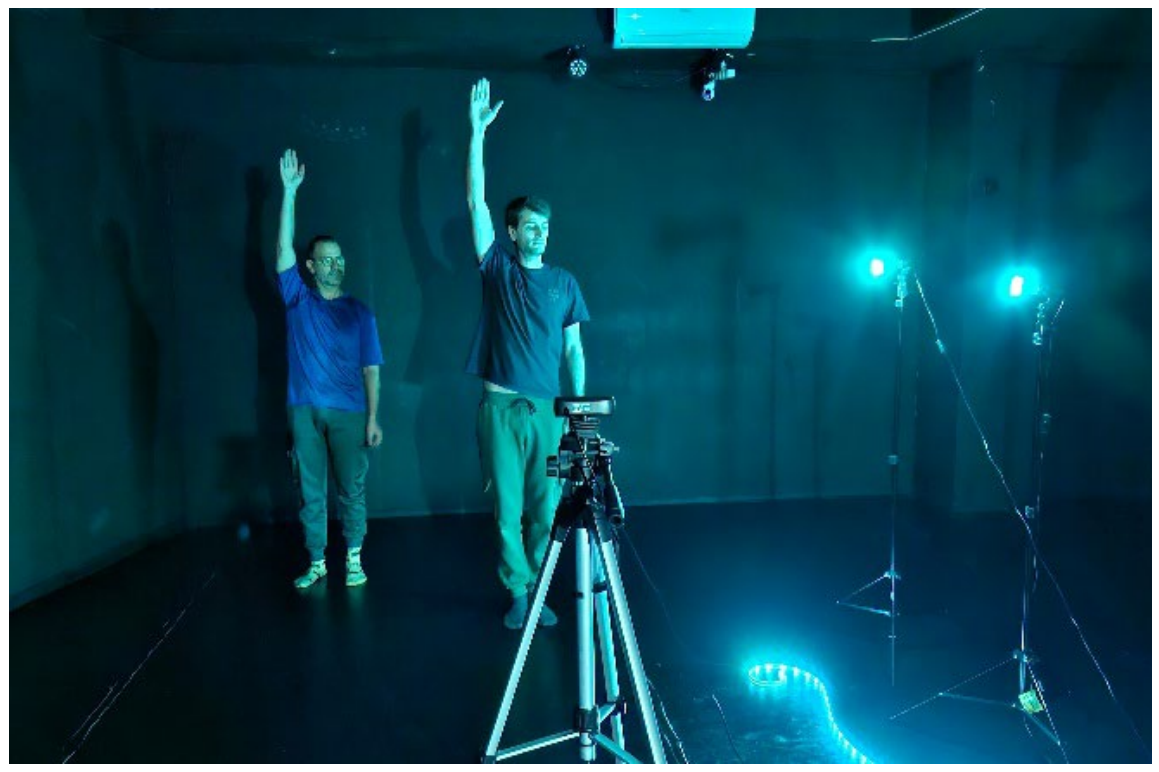

Figure 7: Workshop 2 scene excerpt – gesture triggered lighting changes during a sequence.

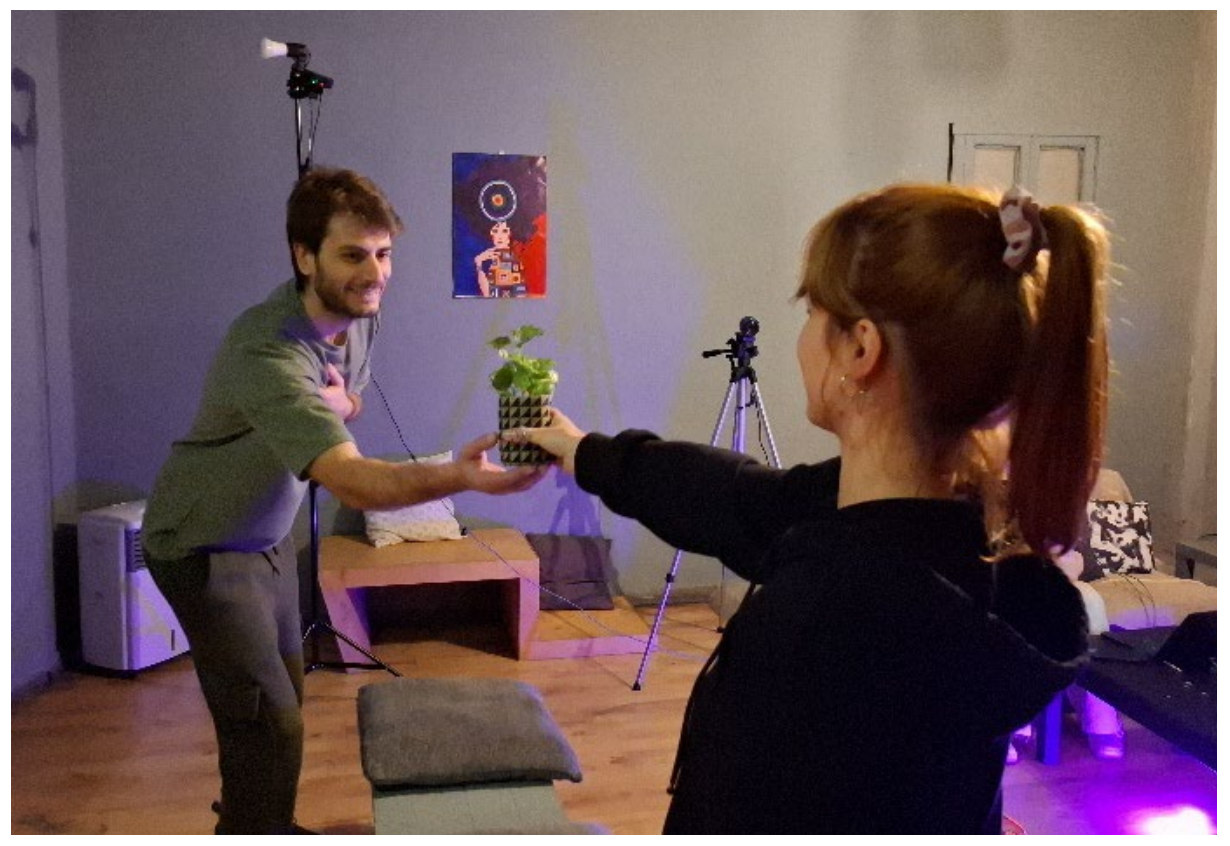

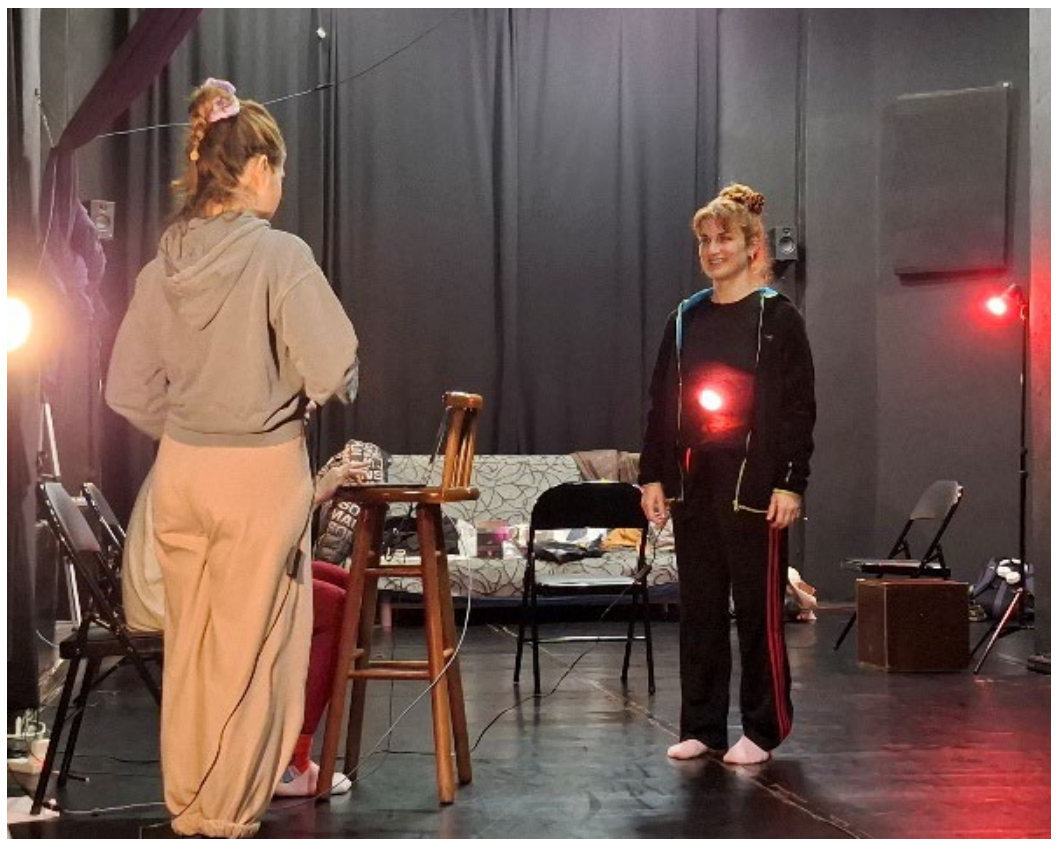

Figure 8: Workshop 5: groups design and test responsive built environments and participatory formats for one audience member.

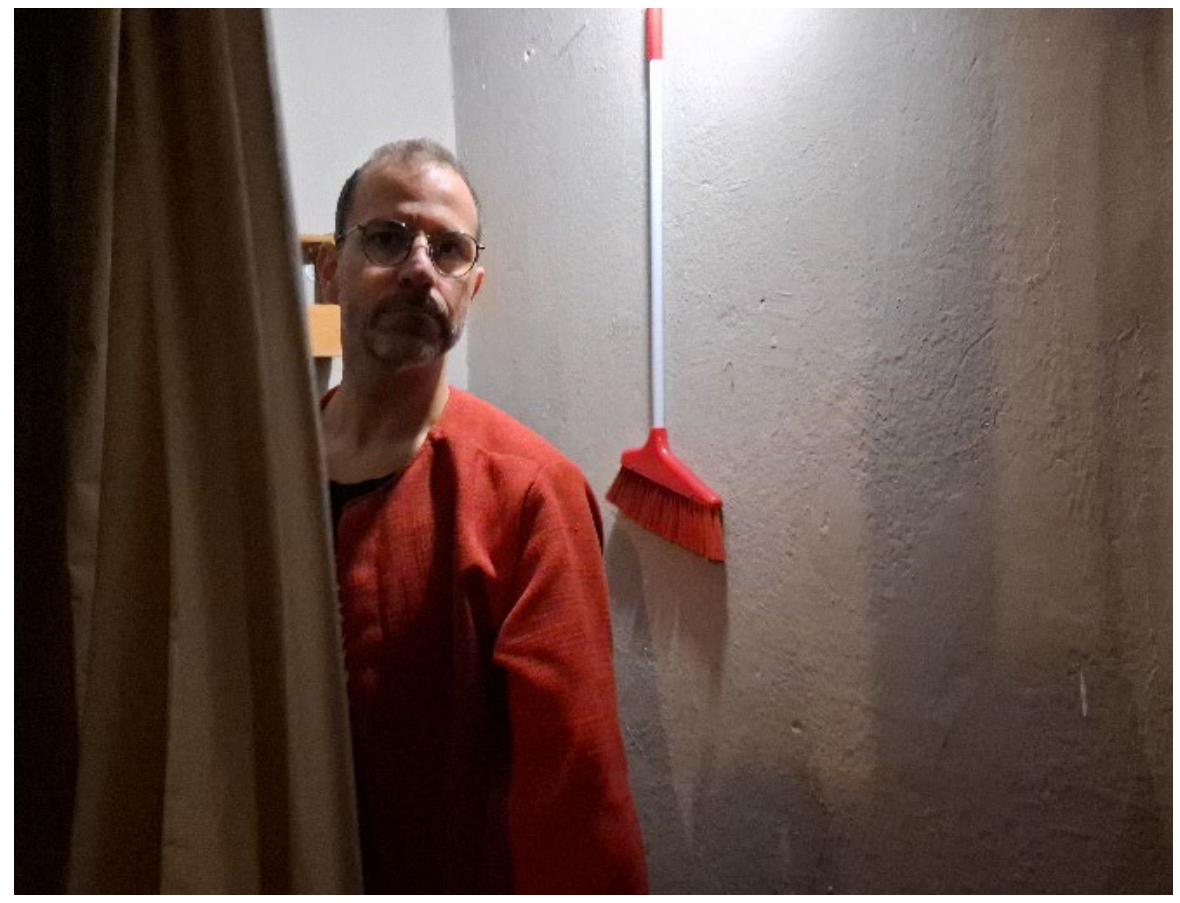

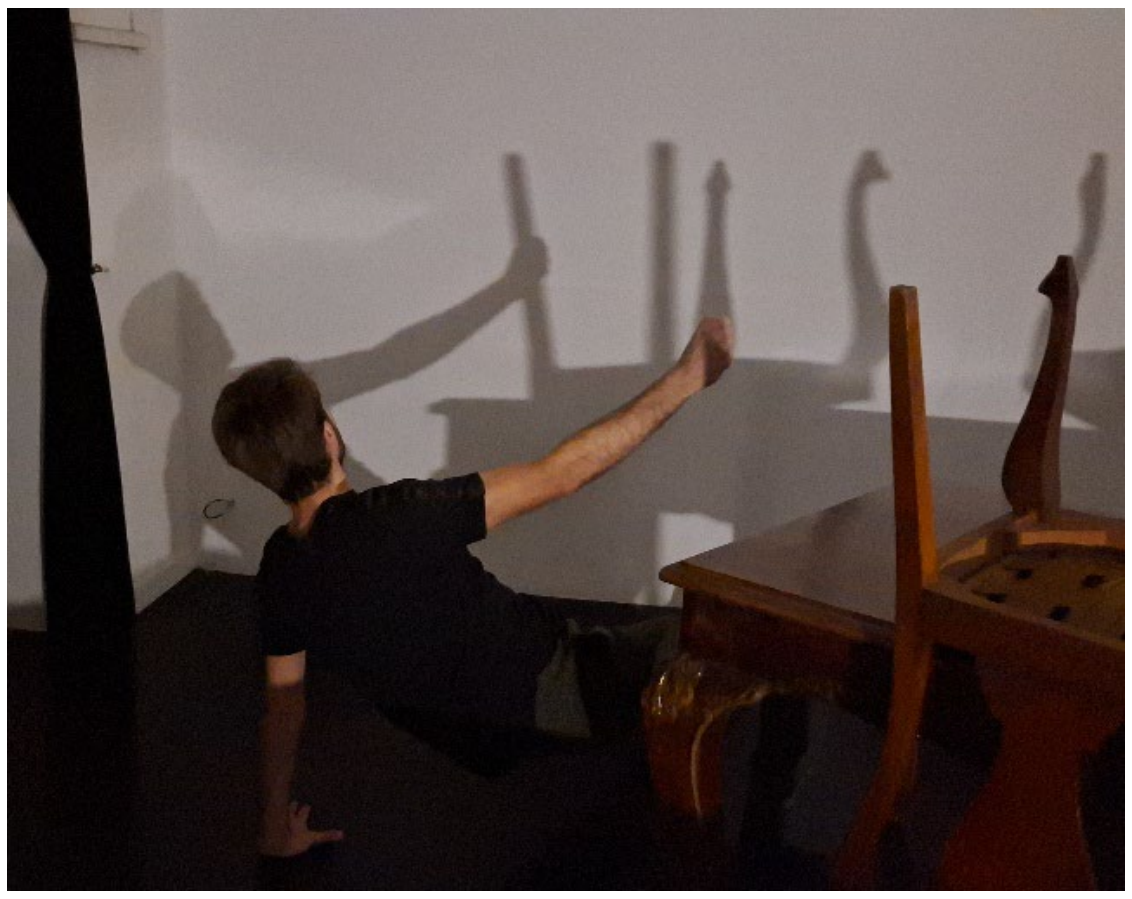

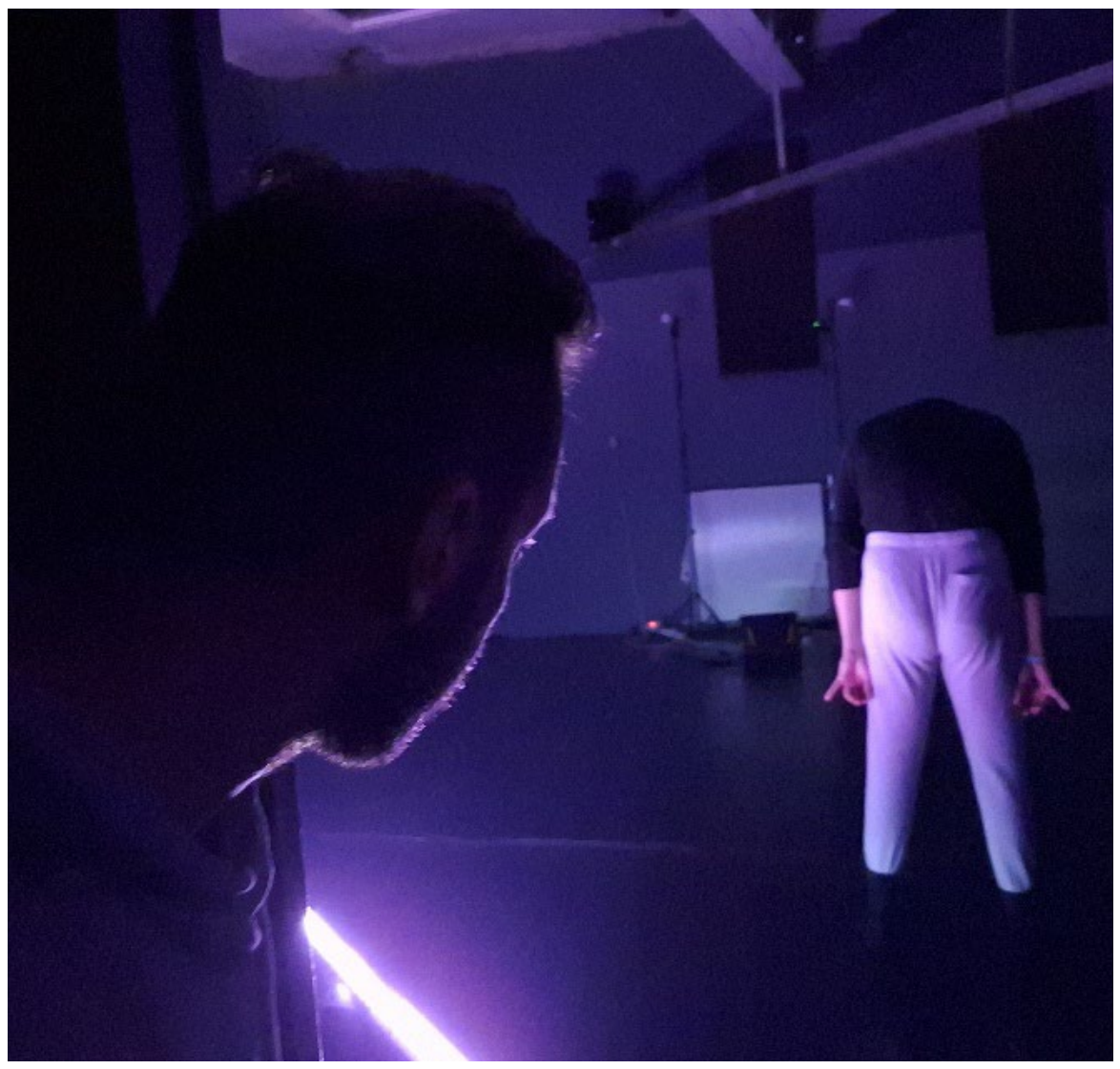
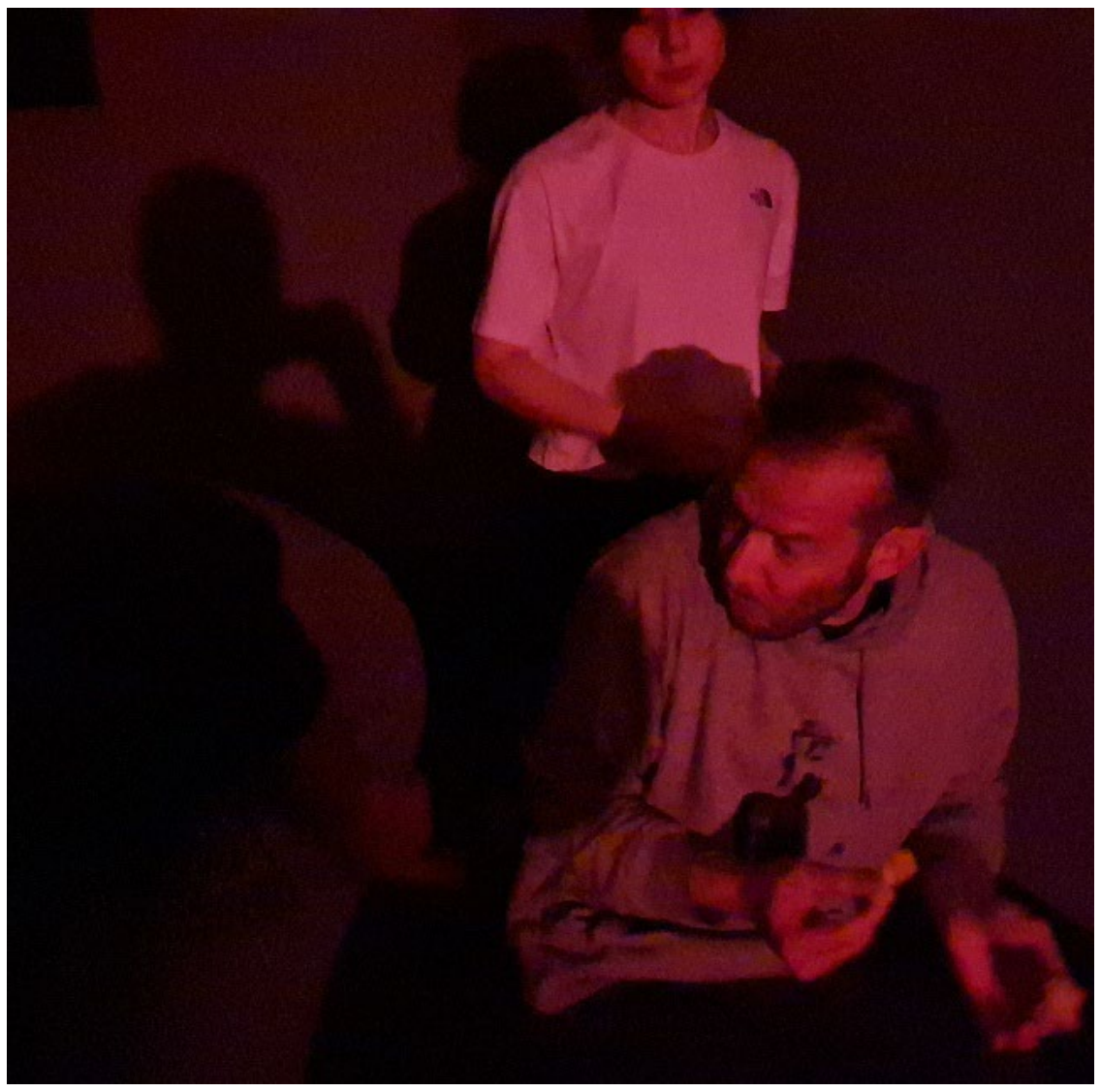

Figure 9: Workshop 6: excerpts from a multi-room immersive scratch performance using a responsive environment in one segment.

## 4. USE IN REHEARSAL AND FUTURE DIRECTION

The workshops suggest that the system's main value lies in making sensing and actuation available as rehearsal materials: simple enough to configure quickly, visible enough to be understood collectively, and flexible enough to be tested and refined through embodied rehearsal. In Workshop 2, participants developed fluency through hands-on trial and error, eventually creating short interactive pieces without direct facilitator intervention. Participants described this as creating a new "vocabulary" between performers and space, where clarity and simplicity made the system more understandable for the group.

As participants became more familiar with the system, its use moved beyond cue-like control toward the design of responsive conditions for others to inhabit. In Workshop 4, groups created bounded architectures for other performers to enter without knowing the mappings in advance, discovering active points and environmental behaviours through physical improvisation. Designers described this as creating small spatial "traps" so that others would "naturally go there". The system therefore supported a shift from composing fixed scenes to shaping conditions for performance to emerge: responsive settings that could guide attention, create atmosphere, and invite others to improvise with the space.

In the later workshops, the system became less of a central object and more of one ingredient within broader immersive composition. In Workshop 5, participants used it alongside movement, sound, objects, text, spatial arrangement, and performer interaction. Participants noted that the technology "did not impose itself" and that "the important thing was the interaction we had." Workshop 6 extended this into a compact performance-making sprint: the group developed a short immersive experience for one spectator across multiple rooms, using the interactive system only where it served a specific dramaturgical purpose. In this sense, the system's usefulness was not that it had to drive the whole performance, but that it could be introduced selectively where responsive behaviour added something compositionally meaningful.

Looking ahead, this rehearsal-oriented setup provides a basis for an agentic form of responsive architecture. In the present system, performers define explicit mappings between sensing inputs and outputs. Future work will extend this approach by treating the room itself as an AI-enabled scenographic participant: an agent that interprets behaviour through a dramaturgical role, selects from a repertoire of spatial actions, and explains its choices in theatre-making terms such as objective, circumstance, and action. This direction retains the commitment to embodied, in-room experimentation, while moving from authoring responsive cues toward directing spatial agents whose behaviour can be shaped through rehearsal, in ways inspired by working with actors [13].

## 5. CONCLUSION

This demo presented a rehearsal-oriented system for integrating bespoke sensing into theatre devising workshops. The aim was not to develop a highly accurate or production-ready interactive system, but to create a setup that could be deployed quickly in different spaces, used by performance-makers without extensive technical expertise, and support experimentation with gesture, position, and speech as triggers for light and sound. Across six workshops, the system was usable enough to support rehearsal-based authoring: participants configured mappings, tested them in situ, and incorporated responsive behaviours into short scenes, responsive environments, and a multi-room experience. This suggests that interactive architectures can be introduced into devising immersive theatre without fully stabilising sensing or requiring a specification-first workflow.

At the same time, the system's bespoke nature and limited calibration introduced fragility. Recognition was not always reliable, and configuration remained approximate. Rather than being fully resolved, this fragility became part of the working conditions of rehearsal. Participants adapted through simple mappings, incremental testing, and by treating the system's occasional unreliability as material for composition. For the purposes of this demo, the main conclusion is that interactive sensing systems can function as compositional tools in rehearsal when they support quick setup, visible logic, and in-room experimentation. Under these conditions, they can be integrated into devising as one ingredient among others, allowing performers to explore responsive environments while ideas are still being formed. Responsive architectures therefore do not need to be fully stable to enter rehearsal, as long as they remain usable within the logics of performance-making.

## ACKNOWLEDGMENTS

This work was supported by the Engineering and Physical Sciences Research Council [grant number EP/T022493/1], Lakeside Arts, Makers of Imaginary Worlds, and the Horizon Centre for Doctoral Training at the University of Nottingham.